\documentclass[fleqn,usenatbib,useAMS]{mnras}

\usepackage{graphicx}	% Including figure files
\usepackage{amsmath}	% Advanced maths commands
\usepackage{amssymb}	% Extra maths symbols
\usepackage{multicol}        % Multi-column entries in tables
\usepackage{bm}		% Bold maths symbols, including upright Greek
\usepackage{pdflscape}	% Landscape pages
\usepackage{csvsimple}

\usepackage[T1]{fontenc}
\usepackage{ae,aecompl}

\usepackage{newtxtext,newtxmath}
\title[Magellanic Cloud pulsars]
{The Thousand-Pulsar-Array programme on MeerKAT VII: Polarisation properties of pulsars in the Magellanic Clouds}
\author[Johnston et al.]
{S.~Johnston$^{1}$\thanks{Email: simon.johnston@csiro.au},
A.~Parthasarathy$^{2}$,
R.~A.~Main$^{2}$,
J.~P.~Ridley$^{3}$,
B.~S.~Koribalski$^{1}$,\newauthor
M.~Bailes$^{4,5}$,
S.~J.~Buchner$^{6}$,
M.~Geyer$^{6}$,
A.~Karastergiou$^{7,8}$,
M.~J.~Keith$^{9}$,\newauthor
M.~Kramer$^{2}$,
M.~Serylak$^{10,11}$,
R.~M.~Shannon$^{4,5}$,
R.~Spiewak$^{9,4,5}$,\newauthor
V.~Venkatraman Krishnan$^{2}$
\\
$^1$Australia Telescope National Facility, CSIRO Space and Astronomy, PO~Box~76, Epping NSW~1710, Australia\\
$^2$Max-Planck-Institut f{\"u}r Radioastronomie, Auf dem H{\"u}gel 69, D-53121 Bonn, Germany\\
$^3$School of Engineering, Murray State University, Murray, KY 42071, USA\\
$^4$Centre for Astrophysics and Supercomputing, Swinburne University of Technology, Hawthorn, VIC 3122, Australia\\
$^5$ARC Centre of Excellence for Gravitational Wave Discovery (OzGrav)\\
$^6$South African Radio Astronomy Observatory (SARAO), 2 Fir Street, Black River Park, Observatory, Cape Town, 7925, South Africa\\
$^7$Department of Astrophysics, University of Oxford, Denys Wilkinson Building, Keble Road, Oxford OX1 3RH, UK\\
$^8$Department of Physics and Electronics, Rhodes University, PO Box 94, Grahamstown 6140, South Africa\\
$^9$Jodrell Bank Centre for Astrophysics, Department of Physics and Astronomy, University of Manchester, Manchester M13 9PL, UK\\
$^{10}$ SKA Observatory, Jodrell Bank, Lower Withington, Macclesfield, SK11 9FT, United Kingdom\\
$^{11}$Department of Physics and Astronomy, University of the Western Cape, Bellville, Cape Town, 7535, South Africa
}
\date{Last updated; in original form}

\pubyear{2021}

\begin{document}
\label{firstpage}
\pagerange{\pageref{firstpage}--\pageref{lastpage}}
\maketitle

\begin{abstract}
The Magellanic Clouds are the only external galaxies known to host radio pulsars. The dispersion and rotation measures of pulsars in the Clouds can aid in understanding their structure, and studies of the pulsars themselves can point to potential differences between them and their Galactic counterparts. We use the high sensitivity of the MeerKAT telescope to observe 17 pulsars in the Small and Large Magellanic Clouds in addition to five foreground (Galactic) pulsars. We provide polarization profiles for 18 of these pulsars, improved measurements of their dispersion and rotation measures, and derive the mean parallel magnetic field along the lines of sight. The results are broadly in agreement with expectations for the structure and strength of the magnetic field in the Large and Small Magellanic Clouds. The Magellanic Cloud pulsars have profiles which are narrower than expected from the period-width relationship and we show this is due to selection effects in pulsar surveys rather than any intrinsic difference between the population of Galactic and Magellanic objects.
\end{abstract}


%%%%%%%%%%%%%%%%%%%%%%%%%%%%%%%%%%%%%%%%%%%%%%%%%%

%%%%%%%%%%%%%%%%% BODY OF PAPER %%%%%%%%%%%%%%%%%%
\section{Introduction}
More than 3000 radio pulsars have been discovered over the past 50 years, nearly all of which are located within our Galaxy and its associated globular cluster system. Only two external galaxies have had radio pulsars discovered within them; the Small and Large Magellanic Clouds (SMC and LMC hereafter). To date, seven radio pulsars have been discovered within the SMC and 22 within the LMC, all with the Parkes telescope \citep{mhah83,mmh+91,mml+93,kjb+94,ckm+01,mfl+06,rcl+13,tsm+19}. At distances of 62~kpc (SMC; \citealt{gpt+20}) and 50~kpc (LMC; \citealt{pgg+19}) these pulsars are typically a factor of ten further than the pulsars detected in our Galaxy and hence their flux densities are very low.
\begin{table*}
\caption{Information for the observed pulsars. In the second column, SMC denotes pulsars in the Small Magellanic Cloud, LMC pulsars in the Large Magellanic Cloud and G pulsars in our Galaxy. The third column lists the observation time in minutes. Unless stated otherwise, values in columns 6 through 9 are from this work. Values in parentheses denote uncertainties in the last digit.}
\label{tab_results}
\begin{center}
    \begin{tabular}{lrrrrrrll}
    \hline
    \hline
JNAME & ID & T$_{\rm obs}$ & $P$ & log $\dot{E}$ &  DM  & RM & $S_{\rm 1400}$ & $\Delta\nu_d$\\
      & & (min) & (s) & (erg\,s$^{-1}$) & (cm$^{-3}$\,pc) & (rad\,m$^{-2}$) & (mJy) & (MHz)\\
\hline
J0045--7042  & SMC & 37 &  0.632336   &   32.5  &    71.3(3) &  +31(3) & 0.064(4) \\
J0045--7319  & SMC & 10 &  0.926276   & 32.3  &   104.6(4) &  --25(3)  & 0.25(1) & 0.99(8)\\
J0111--7131  & SMC & 37 &  0.688542   & 32.9  &    79.4(5) &     &   0.044(4) & 1.0(3)\\
J0113--7220  & SMC & 10 &  0.325883   & 33.7  &   125.49(2)& +128(2) &    0.189(8) & 0.34(3) \\
J0131--7310  & SMC & 31 &  0.348124   & 33.2  &   205.22(6)& --57(3) &   0.059(4)\\
\\
J0449--7031  & LMC & 30 &  0.479164   & 33.0  &    65.77(7)& +41(3) &   0.056(4)) \\
J0455--6951  & LMC & 27 &  0.320423   & 34.0  &    94.70(2)& --26(1)  &  0.176(5) & 0.30(2)\\
J0456--7031  & LMC & 14 &  0.800132   & 33.4  &   100.3(3)$^c$&       &  0.013(7)\\
J0502--6617  & LMC &  5 &  0.691251   & 33.4  &    68.9(3)$^c$  \\
J0519--6932  & LMC & 20 &  0.263212   & 33.1  &   118.86(2) &  +85(6) & 0.13(1) & 0.21(5)\\
J0522--6847  & LMC & 37 &  0.674532   & 33.3  &   126.2(2)  & --123(5) &   0.083(3) \\
J0529--6652  & LMC & 22 &  0.975737   & 32.8  &   103.31(7) & +9(2) &  0.213(6) & 0.12(3) \\
J0532--6639  & LMC & 36 &  0.642743   & 32.9  &    69.2(2) & +54(10) &   0.042(4) \\
J0534--6703  & LMC & 31 &  1.817565   & 33.4  &    95.3(1) & --35(1) &  0.116(5) & 0.24(6)\\
J0540--6919  & LMC &    &  0.050570   & 38.1  &   147.0(1)$^a$  & --246(1)$^a$ &  0.10(3)$^a$ & 0.00036$^a$\\
J0543--6851  & LMC & 42 &  0.708954   & 32.6  &   134.9(6) &    &   0.087(4) \\
J0555--7056  & LMC & 44 &  0.827838   & 32.6  &    72.9(1) &  +26(6) &   0.058(4) & 1.6(2)\\
\\
J0133--6957  & G   &  5 &  0.463474   & 31.6  &    22.95(1)   &    20(5)  &   0.18(1) & 4.6(4) \\
J0457--6337  & G   & 43 &  2.497012   & 29.7  &    27.5(1.0)$^c$   &   14(20)$^b$  &0.030(5) \\
J0511--6508  & G   & 24 &  0.322062   & 32.3  &    25.68(9)   &   27(6)  &    0.31(1) & 4.8(6)\\
J0536--7543  & G   & 21 &  1.245856   & 31.0  &    18.60(2) &   26.1(1)&   8.70(1) & 3.0(3)\\
J0540--7125  & G   & 27 &  1.286015   & 31.1  &    29.97(8) &   40(2)  &   0.247(5) & 2.2(2) \\
\hline
    \end{tabular}
\end{center}
References: $^a$ \citet{gsa+21},
$^b$ \citet{hmvd18},
$^c$ \citet{mfl+06}.
\end{table*}

In our own Galaxy, pulsars make excellent probes of the magneto-ionic interstellar medium (ISM). A pulsar's dispersion measure (DM) is proportional to the integrated free-electron density along the line of sight and can be used to estimate the pulsar's distance (e.g. \citealt{sch12,ymw17}). The rotation measure (RM), combined with the DM, allows the mean parallel magnetic field to be derived (e.g. \citealt{sbg+19}), leading to models of the global magnetic field structure of the Galaxy (e.g. \citealt{njkk08,hmvd18}). Finally, the dynamic spectrum (e.g. \citealt{bmg+10}) and long-term flux variability (e.g. \citealt{kdj+21}) reveal the turbulence and inhomogeneous nature of the ISM. A similar approach has been attempted in the LMC and SMC in spite of the relatively small number of sight-lines available. \citet{mfl+06} looked at the electron density distribution in the LMC, later updated by \citet{ymw17}. Use of extra-galactic RMs seen through the LMC \citep{mmg+12} and SMC \citep{mgs+08} allowed a model of their global magnetic fields to be constructed.

The radio luminosity of a pulsar at 1400~MHz, crudely defined as $L = S_{\rm 1400}~d^2$ where $S_{\rm 1400}$ is the flux density at 1400~MHz and $d$ the distance in kpc, varies by several orders of magnitude across the population with only a mild dependence on its spin parameters (e.g. \citealt{szm+14}). Although the distance determination in our Galaxy is subject to considerable uncertainty, the known pulsars in the Magellanic Clouds are at large distances and hence are amongst the most luminous pulsars known. Examination of the Magellanic pulsars can therefore indicate whether luminous pulsars are different from the less-luminous ones \citep{rcl+13} and/or whether distant pulsars have different characteristics to nearby ones.

In this paper we present observations from 22 pulsars, 17 of which are in the Magellanic Clouds. The observations and data analysis are described in Section~2 and Section~3 presents the polarization profiles and details on the individual pulsars. In Section~4 we discuss the implications of the results for the Magellanic Clouds and for the pulsar population.

\section{Source selection, observations and analysis}
For the purposes of pulsar observing, the MeerKAT telescope forms a tied-array beam on the sky from the coherent addition of its 64 individual antennas. The tied array beam has a full-width half-maximum of some 6 arcsec at 1.4~GHz and hence the position of the pulsar needs to be known to an accuracy of a few arcsec. Of the 29 radio pulsars in the LMC/SMC, only 17 meet this criteria and are thus suitable for MeerKAT observations. In addition to these pulsars we observed a further 5 pulsars which are in the direction of the Magellanic Clouds, but located within our own Galaxy. Table~\ref{tab_results} lists the observational details of the 22 pulsars observed. Column 2 denotes the location of the pulsars. Column 3 gives the total observing time, column 4 and 5 gives the pulsar spin period ($P$) and spin-down energy ($\dot{E}$) respectively.

Observations were conducted as part of the Thousand Pulsar Array (TPA) program, itself part of the larger MeerTime project. Details of the project as a whole and the observational setup can be found in \citet{mtime} and \citet{tpa20}. In brief, we used the observational band from 896 to 1671~MHz with 928 frequency channels. The data are folded into sub-integrations each of length 8~s for the duration of the observation and there are 1024 phase bins per pulse period. Polarization calibration is carried out using the procedures described in detail in \citet{sjk+21}. Data coming from the telescope have units of digitiser counts and we need to determine a multiplier to convert the units into mJy. To do this we assume that a single MeerKAT antenna has a system equivalent flux density of 390~Jy. We then use the number of antennas employed in conjunction with the bandwidth and observing time to compute the expected rms (in mJy) in a given frequency channel. The data are then scaled appropriately so that the actual rms matches the expected rms. Flux densities computed in this way are listed as $S_{\rm 1400}$ in column 8 of Table~\ref{tab_results}. All these operations are carried out by the processing pipeline \textsc{meerpipe}\footnote{https://github.com/aparthas3112/meerpipe} which produces RFI-excised, polarisation- and flux-calibrated output products in \textsc{psrfits} format \citep{hvm04}. A more complete description of the pipeline is reported in \citet{pbs21}.
\begin{figure*}
\begin{center}
\begin{tabular}{ccc}
\includegraphics[width=5cm,angle=0]{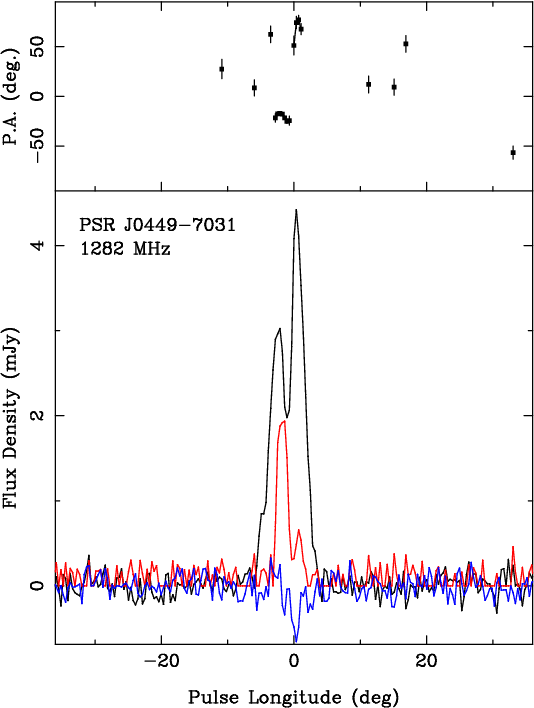} &
\includegraphics[width=5cm,angle=0]{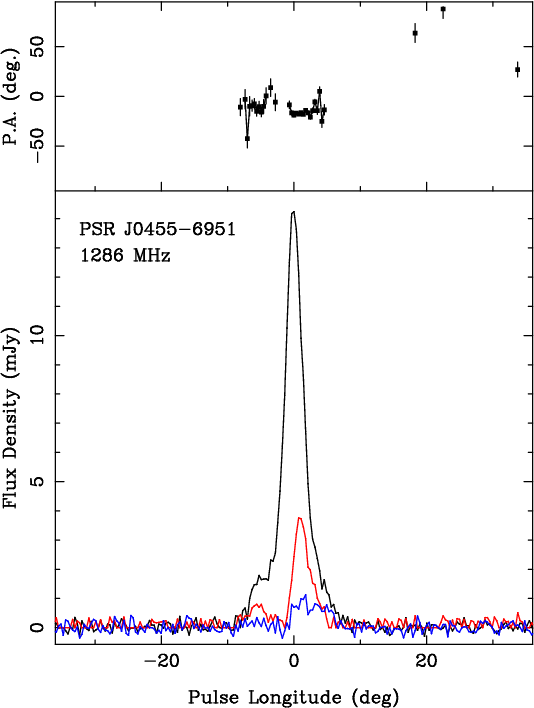} &
\includegraphics[width=5cm,angle=0]{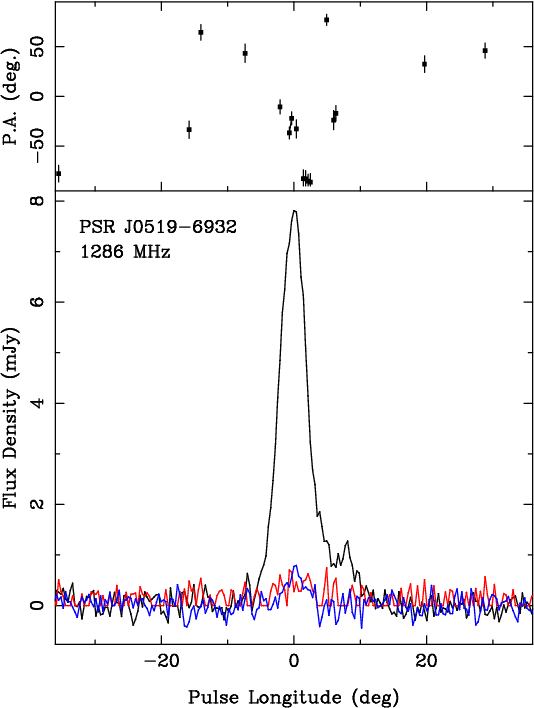} \\
\includegraphics[width=5cm,angle=0]{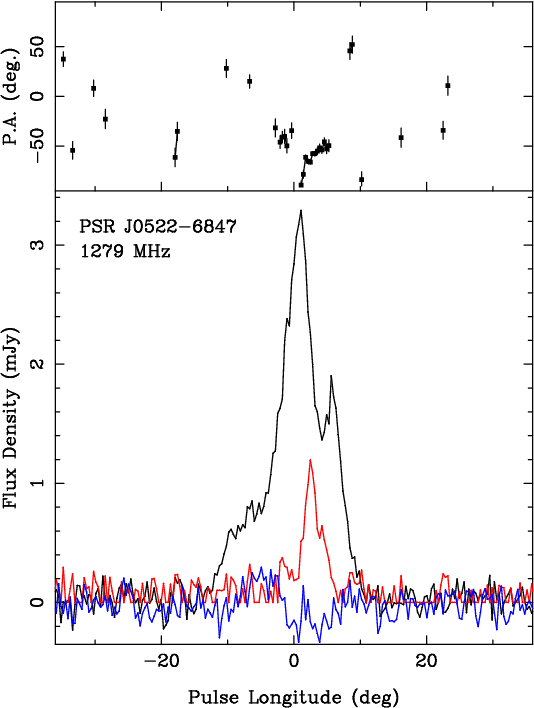} &
\includegraphics[width=5cm,angle=0]{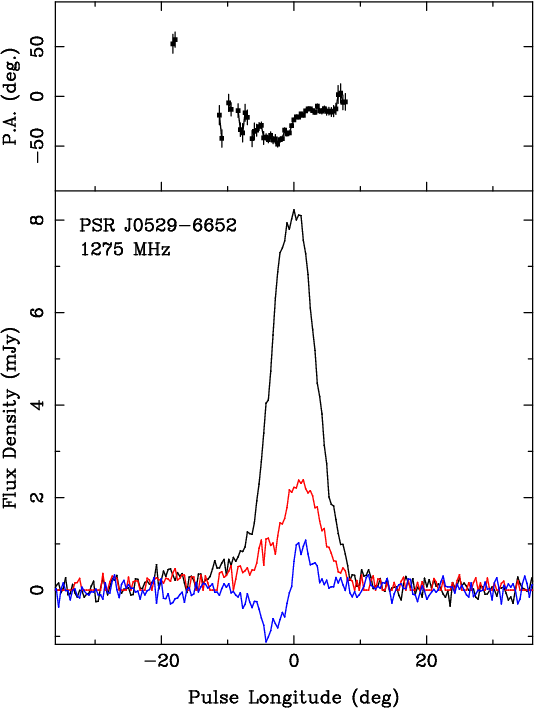} &
\includegraphics[width=5cm,angle=0]{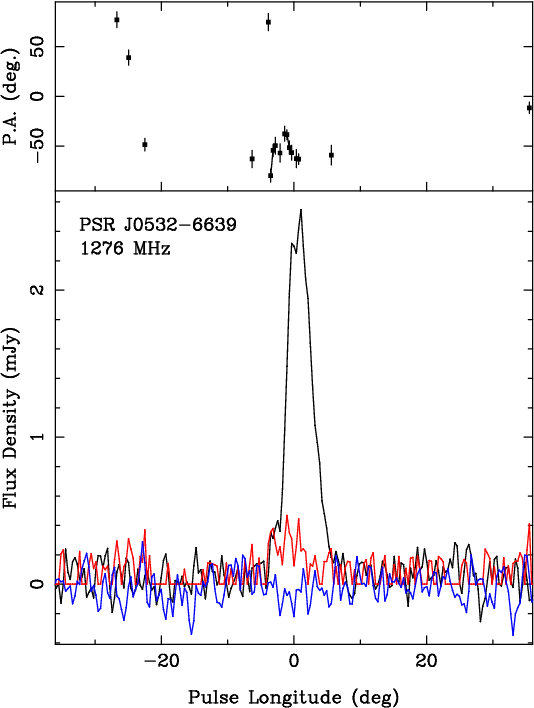} \\
\includegraphics[width=5cm,angle=0]{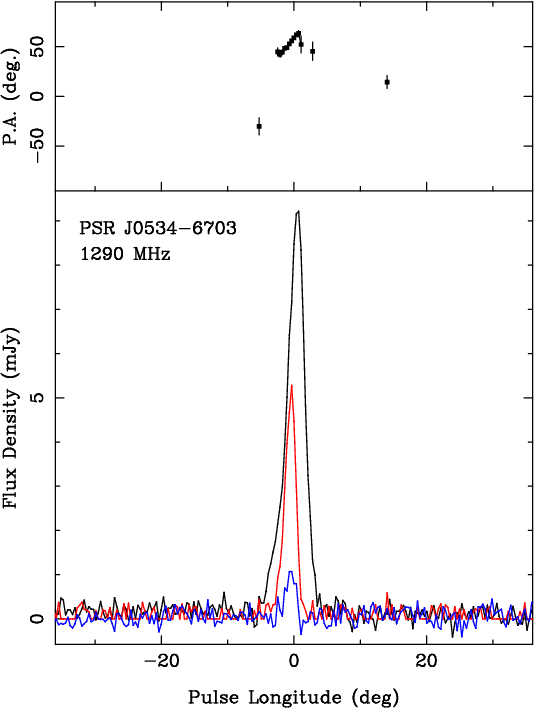} &
\includegraphics[width=5cm,angle=0]{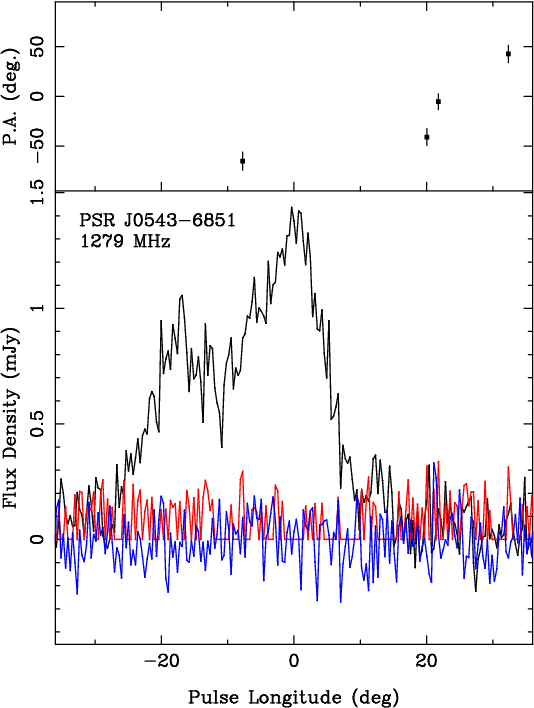} &
\includegraphics[width=5cm,angle=0]{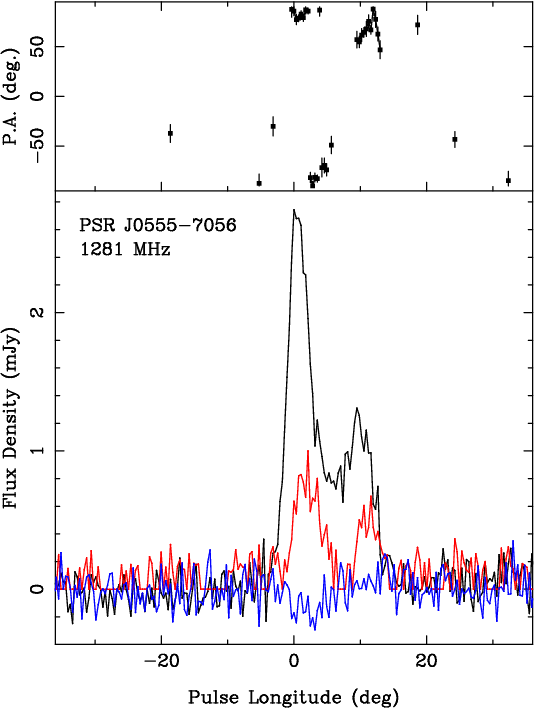} \\
\end{tabular}
\end{center}
\caption{Polarisation profiles for the LMC pulsars. In the lower panels, the black line denotes Stokes I, the red trace shows the linear polarisation and the blue trace the circular polarisation. Left-hand circular polarisation is defined to be positive. The top panel shows the position angle of the linear polarisation, corrected to infinite frequency using the RM listed in Table~\ref{tab_results}. Position angles are only plotted when the linear polarisation exceeds 3 sigma. The zero point of pulse longitude is set to the peak of the total intensity profile.}
\label{figlmc}
\end{figure*}

The dispersion measure (DM) is computed as follows. Using the pulsar ephemeris, we sum all the data in frequency and time to produce a noiseless, smoothed template. We also produce data summed in time and reduced to 32 frequency channels. The {\sc psrchive} \citep{hvm04} routine {\sc pat} is then used in conjunction with the template to produce a time-of-arrival of the profile in each of the frequency channels. These arrivals times are then passed into {\sc tempo2} \citep{hem06} and a fit to the DM is made.  The rotation measure (RM) is also computed from the same data set (i.e. time summed and with 32 frequency channels) via the routine {\sc rmfit}. The derived DMs, RMs and their uncertainties are given in columns 6 and 7 of Table~\ref{tab_results}.

We attempted to measure the scintillation bandwidth, $\Delta\nu_d$, for the pulsars at 1~GHz. For the foreground pulsars with $\Delta\nu_d$ greater than the channel bandwidth this was computed in the standard way via the auto-correlation function (ACF) of the dynamic spectrum (e.g. \citealt{jnk98}) over 50~MHz of bandwidth centered at 1~GHz.
For pulsars with scintillation marginally unresolved at 1~GHz, $\Delta\nu_d$ was computed in 200~MHz centered on 1.4~GHz, and the value scaled to 1~GHz with a --4 index.
For the remaining pulsars, $\Delta\nu_d$ was estimated via $m^2 \Delta\nu$ where $m$ is the modulation index and $\Delta\nu$ the channel bandwidth (e.g. \citealt{kcw+18}). Measurements were made across the observing bandwidth and a power-law fit was obtained to yield the value at 1~GHz. Conservatively the uncertainty was set to 25\% of the value \citep{kcw+18}. Where obtainable, $\Delta\nu_d$ is listed in column 9 of Table~\ref{tab_results}. A more complete description of the methodology will be given in an upcoming paper on the scintillation properties of the entire TPA pulsar sample.

\section{Pulsar Profiles}
\begin{figure*}
\begin{center}
\begin{tabular}{ccc}
\includegraphics[width=5cm,angle=0]{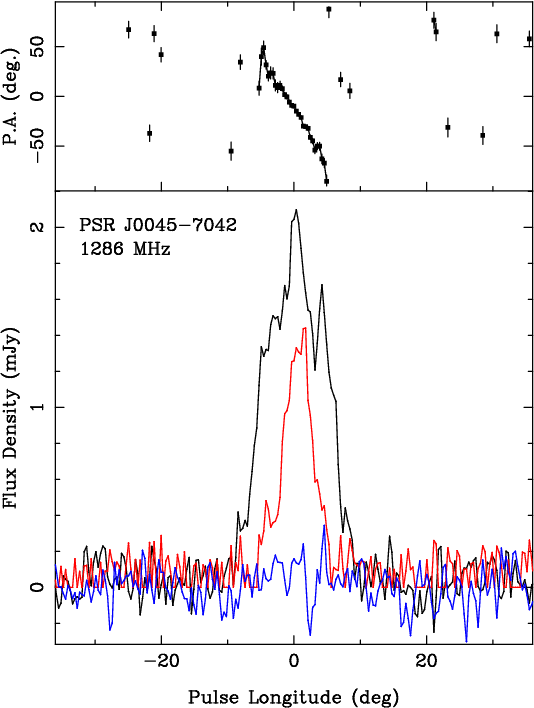} &
\includegraphics[width=5cm,angle=0]{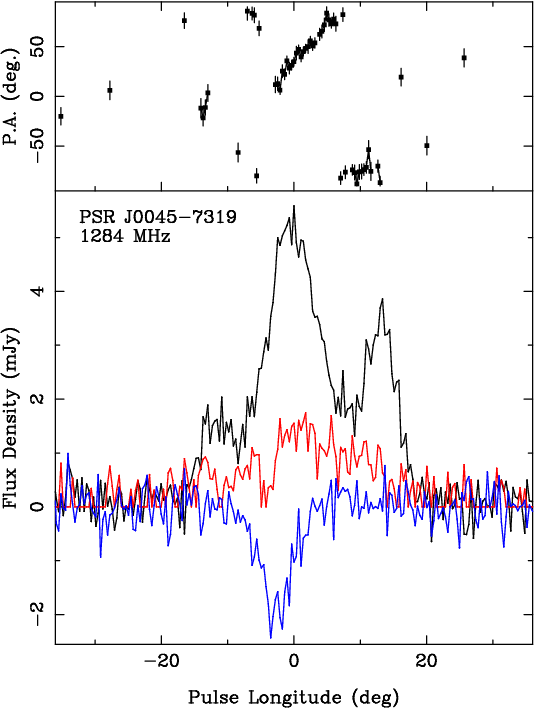} &
\includegraphics[width=5cm,angle=0]{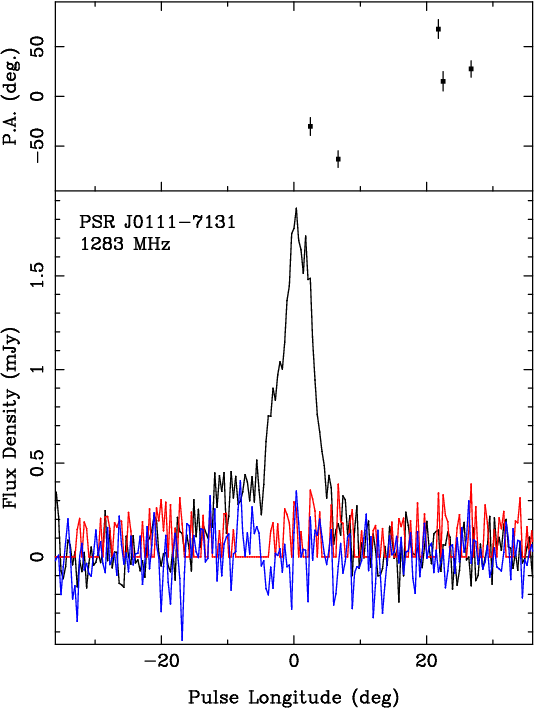} \\
\includegraphics[width=5cm,angle=0]{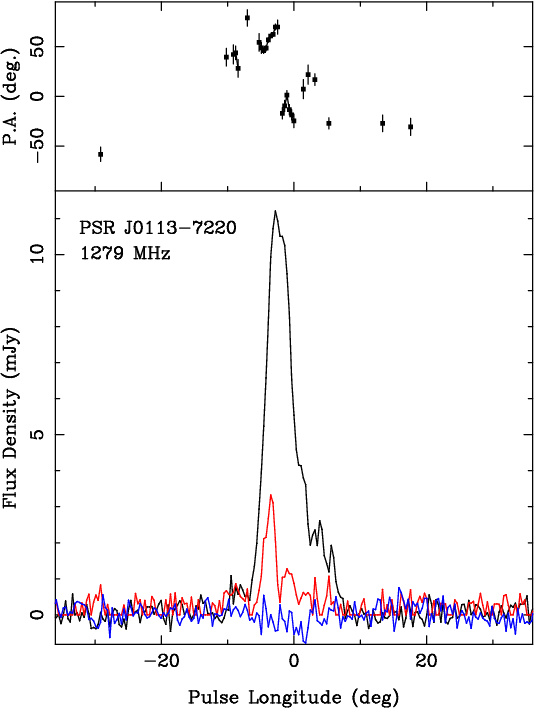} &
\includegraphics[width=5cm,angle=0]{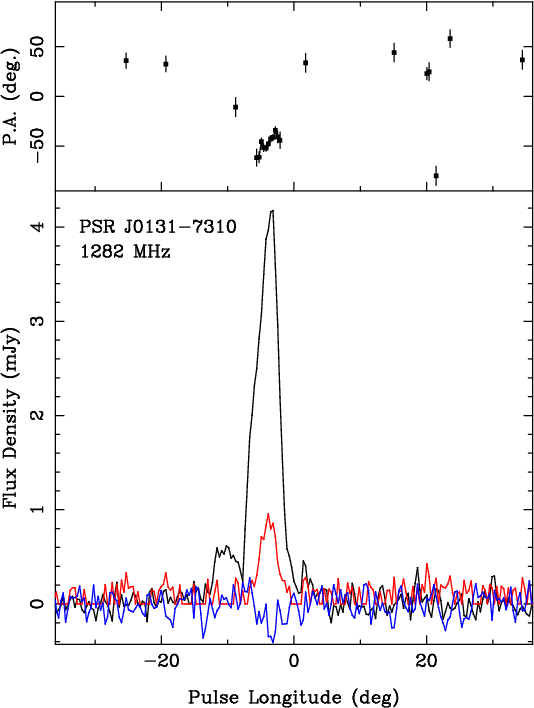} \\
\end{tabular}
\end{center}
\caption{Polarisation profiles for the SMC pulsars. See Figure~\ref{figlmc} for details.}
\label{figsmc}
\end{figure*}

\begin{figure*}
\begin{center}
\begin{tabular}{ccc}
\includegraphics[width=5cm,angle=0]{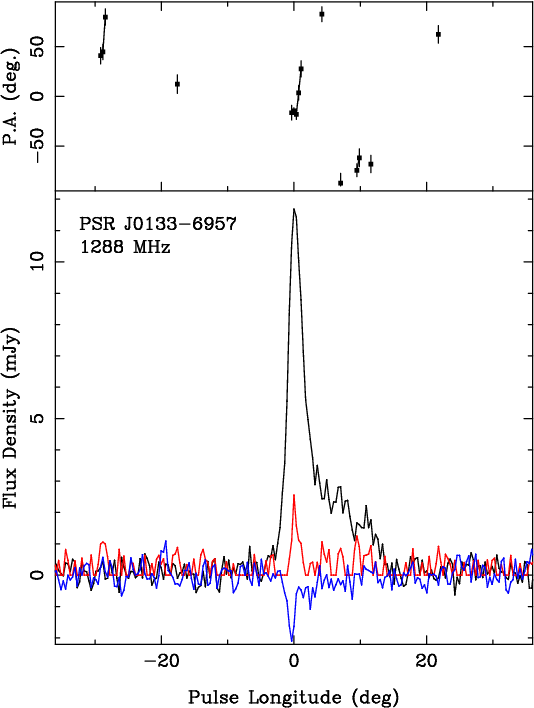} &
\includegraphics[width=5cm,angle=0]{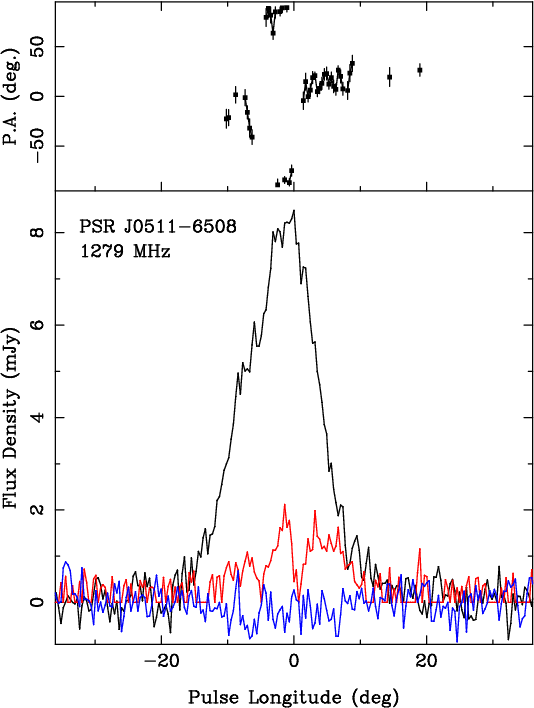} &
\includegraphics[width=5cm,angle=0]{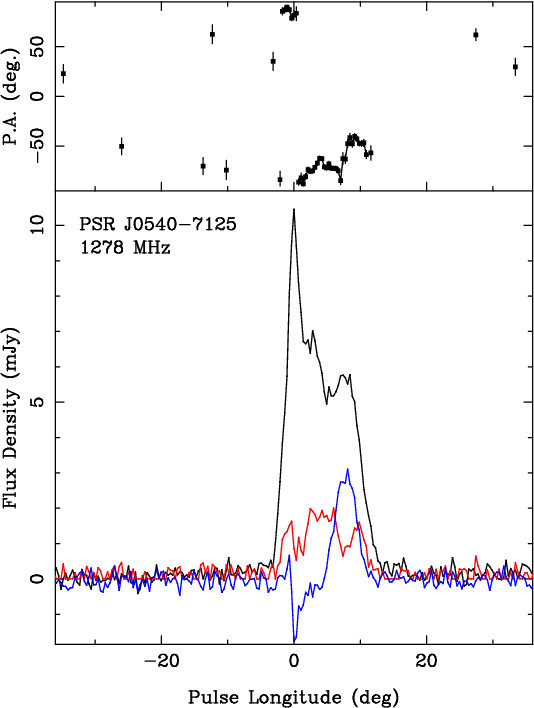} \\
\includegraphics[width=5cm,angle=0]{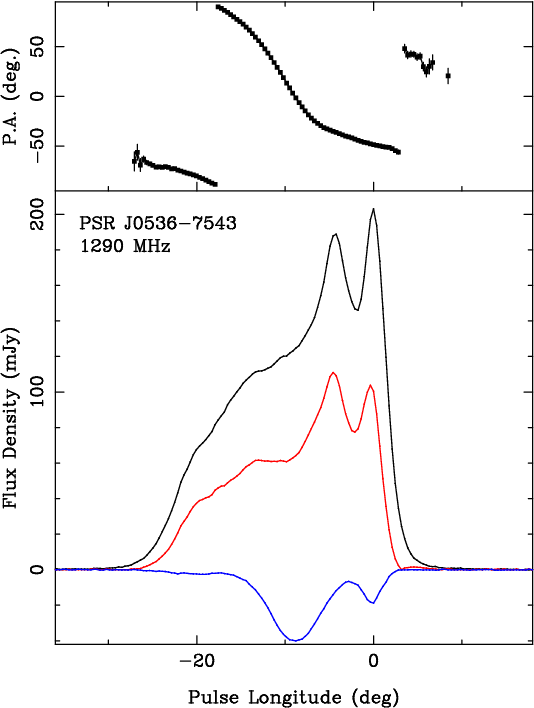} \\
\end{tabular}
\end{center}
\caption{Polarisation profiles for the foreground pulsars. See Figure~\ref{figlmc} for details.}
\label{figmw}
\end{figure*}

The pulsar profiles are shown in Figures~\ref{figlmc} through \ref{figmw}.
Position angles are defined as increasing counter-clockwise on the sky (see \citealt{ew01}) and are corrected to infinite frequency using the RM given in Table~\ref{tab_results}. Circular polarisation in pulsar astronomy uses the IEEE convention for left-hand and right-hand and so in the profiles shown here, left-hand circular polarisation is positive \citep{vmjr10}. The linear polarization is corrected for bias following the prescription in \citet{ew01}. A brief description of the pulsars follows below.

\subsection{LMC pulsars}
The profiles of nine of the LMC pulsars are shown in Figure~\ref{figlmc}. Three pulsars are not shown. PSR~J0502--6617 was not detected (albeit in a short 5~min observation) and PSR~J0456-7031 has only a very low signal-to-noise ratio in a 14~min observation. PSR~J0540--6919 is a young Crab-like pulsar which emits giant pulses \citep{jr03,gsa+21}. The polarisation profile for this pulsar as seen by MeerKAT has been published separately by \citet{gsa+21} and an RM of --246~rad\,m$^{-2}$ was derived.

In general, the profiles of the nine pulsars are narrow, with only PSR~J0543--6851 having a width greater than 20\degr. Three of the pulsars show clear evidence for a double-peaked structure, five have a single component,  and PSR~J0522--6847 appears to have at least three components. The overall polarization fraction is low, typical also for the Galactic pulsars with similar spin-down energies \citep{jk18}. The low polarization fraction coupled with the narrow profiles makes it hard to discern the swing of position angle and no conclusions can be drawn about the spin-axis orientation of these pulsars. Of the pulsars not observed, those presented in \citet{rcl+13} are all single component profiles apart from PSR~J0532--69 which is a narrow double. Polarization information for those pulsars is not available.

Our observations show that PSR~J0529--6652 undergoes episodes of nulling, where the radio emission appears to cease. In the 22~min of observing, it is in the on-state for some 8 minutes and nulling for the remainder. The null states are several minutes in duration.

\subsection{SMC pulsars}
The profiles of the SMC pulsars are shown in Figure~\ref{figsmc}. Four of the five pulsars have profiles with narrow widths of less than 20\degr. Polarization fractions are generally weak, although PSR~J0045--7042 has moderate linear polarization. We have derived RMs for four of the five pulsars. Two had previous measurements and our values are consistent with these but have much smaller uncertainties. We see evidence for wide-band diffractive scintillation at our observing frequency, with flux densities changing by a factor of a few between observations (see Section~4.3).

PSR~J0045--7319 is a pulsar in a binary orbit with a B star companion \citep{kjb+94,bbs+95}. We have 12 observations of this pulsar made over a period of two years. Each observation is typically 6~mins in duration. In seven of these observations the pulsar is either not detected or extremely weak with a peak flux density below 2~mJy. In contrast, two observations have peak flux densities of 20~mJy. These variations are likely due to diffractive scintillation with a bandwidth comparable to the observing bandwidth. We also note that the pulsar has a steep spectral index and is much brighter at 660~MHz than at 1400~MHz. Previous polarization observations of the pulsar by \citet{cmk01} taken at 660~MHz show a low level of linear polarization, relatively high circular polarization and a poorly constrained RM. Their low time-resolution indicates a simple Gaussian profile with a flat swing of PA. Our data at 1300~MHz show the profile has developed into a triple structure. The circular polarization is high, but the swing of position angle is very steep with indications of an orthogonal jump between the first two components. 

\subsection{Foreground pulsars}
The profiles of four foreground pulsars are shown in Figure~\ref{figmw}. Although PSR~J0457--6337 was observed, its profile has a low signal-to-noise ratio and is not shown here. Its flux density of 0.02~mJy is considerably lower than quoted by \citet{mfl+06} and likely the pulsar was in a low scintillation state in our observation. The profiles show a variety of morphologies, including one with three components (PSR~J0540--7125), a trailing-edge dominant profile (PSR~J0536--7543) and a wide single (PSR~J0511--6508). For PSR~J0511--6508 we have multiple, short observations. The flux density varies from 0.05 to 0.9~mJy due to the effects of diffractive scintillation and we quote the mean value in the table.

\section{Discussion}
\subsection{The Magellanic Clouds}
A comprehensive survey of the polarization properties and magnetic field structure of the LMC was carried out by \citet{mmg+12}. The RMs of extragalactic sources seen through the LMC are potentially subject to several screens. The foreground, Milky Way, contribution to the RM is +28~rad\,m$^{-2}$. The LMC has a diffuse halo which (on the near side of the disk) contributes --10~rad\,m$^{-2}$ to the RM. Finally the body of the LMC appears to have an azimuthal magnetic field structure with a strength of 1\,$\mu$G in addition to the presence of clumps, magnetic filaments, supernova remnants on smaller scales \citep{ghs+05,mmg+12}. To compute the RM contribution from the LMC, which we denote RM$_c$, we subtracted 28~rad\,m$^{-2}$ from the observed RM values. Meanwhile, \citet{ymw17} derived the contribution to the DM of the Galactic foreground material in the direction of the LMC pulsars. We use their Table~13 to compute DM$_c$, the DM contribution of the LMC to the overall pulsar DM. Table~\ref{tab_blmc} lists DM$_c$ and RM$_c$ to the LMC pulsars given in Table~\ref{tab_results}. We also compute the inferred magnetic field parallel to the line of sight in $\mu$G via
\begin{equation}
\label{eqn:bpar}
    B_{||} = 1.232 \,\, \frac{{\rm RM_c}}{{\rm DM_c}}
\end{equation}
with DM$_c$ and RM$_c$ in conventional units. Table~\ref{tab_blmc} shows a general trend that as the DM increases the RM becomes more negative. Four of the sight-lines have inferred $B_{||}$ significantly higher than the canonical 1~$\mu$G.
\begin{table}
\caption{DM and RM contributions from the LMC to 12 pulsars, along with the derived value of B$_{||}$ from Equation~\ref{eqn:bpar}.}
\label{tab_blmc}
\begin{center}
    \begin{tabular}{lrrrrrrr}
    \hline
    \hline
JNAME &  DM$_c$  & RM$_c$ & B$_{||}$\\
      & (cm$^{-3}$\,pc) & (rad~\,m$^{-2}$) & ($\mu$G) \\
\hline
J0449--7031  & 16 & +13 &   +1.0 \\
J0455--6951  & 44 & --54 & --1.2  \\
J0456--7031  & 51  \\
J0502--6617  & 19 \\
J0519--6932  & 62 & +57 & +1.1 \\
J0522--6847  & 69 & --151 &   --2.7  \\
J0529--6652  & 45 & --19  &   --0.5  \\
J0532--6639  & 11 & +24 & +2.7  \\
J0534--6703  & 35 & --63 &  --2.2  \\
J0540--6919  & 84 & --274 & --4.0  \\
J0543--6851  & 71 \\
J0555--7056  &  6 &  --2 &  --0.4  \\
\hline
    \end{tabular}
\end{center}
\end{table}

\begin{figure*}
\begin{center}
\begin{tabular}{c}
\includegraphics[width=12cm,angle=0]{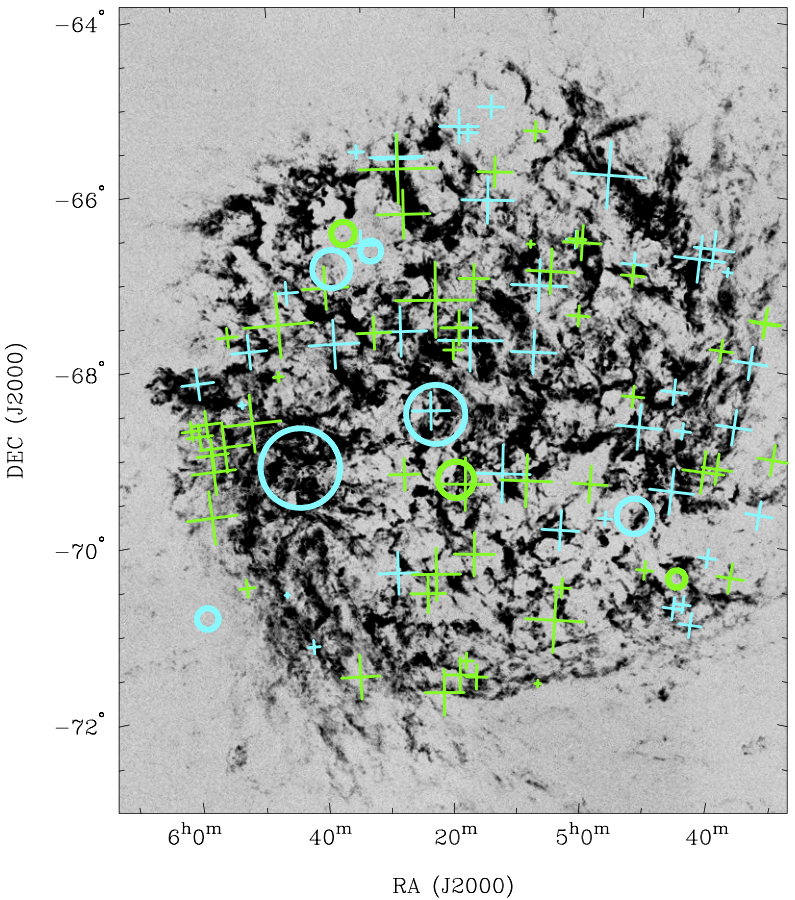} \\
\end{tabular}
\end{center}
\caption{Greyscale image of the LMC, taken from \citet{ksd+98}, with the intensity proportional to the peak flux density in the H{\sc i} line. Crosses mark the location of the 93 extragalactic sources used by \citet{mmg+12}. Circles mark the location of the nine pulsars with measured RMs. Green denotes sight-lines with positive RM, light blue sight-lines with negative RM. The area covered by the symbols is proportional to the RM.}
\label{figbig}
\end{figure*}

In Figure~\ref{figbig} we show an image of the LMC taken from \citet{ksd+98}. Superposed we show the pulsars (circles) from this work and the extragalactic sources (crosses) from \citet{mmg+12}. The colours denote objects with net positive (green) and negative (light blue) values of RM (after subtracting the Galactic foreground contribution). The size of the symbols is proportional to the RM. As summarised by \citet{mmg+12}, the RM values towards extragalactic sources behind the south-eastern part of the LMC are predominantly positive. PSR~J0540--6919, the pulsar with the largest negative RM is located here. However, the very high negative RM and relatively high DM of this pulsar are almost certainly due to electron density content and the magnetic field of its parent plerion and supernova remnant \citep{bmb+14}. This is discussed further in \citet{gsa+21}. Five of our pulsars lie in regions where the extragalactic RMs are close to zero. These pulsars also possess RMs. Finally two of our pulsars, one with negative and one with positive RM, lie close to the centroid of the LMC where the extragalactic RMs also have mixed sign. In summary, with a limited number of sight-lines to pulsars it is difficult to draw conclusions, but the results are broadly in line with the expectations of \citet{mmg+12} although it is clear from some high values of $B_{||}$ that structures local to the pulsar also contribute to the RM and DM.

In the case of the SMC, \citet{mgs+08} estimate that the foreground Galactic material causes an RM of between 0 and $\sim50$~rad\,m$^{-2}$. We have only one foreground pulsar with an RM of 20~rad\,m$^{-2}$. \citet{ymw17} derive a DM of $\sim$30~cm$^{-3}$\,pc for the Galactic foreground. In Table~\ref{tab_bsmc} we list DM$_c$ and RM$_c$ for the five SMC pulsars corrected for the foreground assuming a DM contribution of 30~pc\,cm$^{-3}$) and an RM contribution of +30rad\,m$^{2}$, along with the inferred magnetic field.
\begin{table}
\caption{DM and RM contributions from the SMC to five pulsars, along with the derived value of B$_{||}$ from Equation~\ref{eqn:bpar}.}
\label{tab_bsmc}
\begin{center}
    \begin{tabular}{lrrrrrrr}
    \hline
    \hline
JNAME &  DM$_c$  & RM$_c$ & B$_{||}$\\
      & (cm$^{-3}$\,pc) & (rad~m$^{-2}$) & ($\mu$G) \\
\hline
J0045--7042  & 41 & 1 & 0.1 \\
J0045--7319  & 75 & --55 & --0.9 \\
J0111--7131  & 50 \\
J0113--7220  & 95 & +98 & +1.3 \\
J0131--7310  & 175 & --87 & --0.6 \\
\hline
    \end{tabular}
\end{center}
\end{table}
\citet{mgs+08} examined ten extragalactic sources seen behind the SMC; nine of the ten had negative RMs with a median value of $-75$~rad\,m$^{-2}$. From Table~\ref{tab_bsmc}, for the four pulsars two are negative, one consistent with zero and one (PSR~J0113--7220) positive. Examination of the image of the SMC from \citet{dgma10} shows that PSR~J0113--7220 lies outside the area containing the bulk of the radio continuum emission and there is no immediate explanation for its positive RM. The inferred magnetic fields are in-line with expectations for the SMC.

\subsection{The pulsars}
The SMC and LMC provide us with a set of 29 pulsars at distances beyond 50~kpc. The question arises as to whether there is any difference in the morphology of the profiles between these high-luminosity pulsars and pulsars at a distance of a few kpc. In the quasi-standard picture of pulsars, the radio emission arises at a low altitude above the polar-cap, the region bounded by the open field lines. In some models, emission arises from a central region (`core')  and/or an annulus towards the outside of the beam (`conal') \citep{ran90,kj07}. Emission is seen only when the impact angle comes close to one the magnetic poles. Core emission has a steep spectral index and dominates at low frequencies, whereas conal emission has a much flatter spectral index and hence becomes dominant at higher frequencies. At 1.4~GHz where the majority of the MC pulsars have been found, a mixture of core and cones are seen. In this picture, the distance to the pulsar is not relevant and MC pulsars should have the same properties as Galactic pulsars. A contrasting model postulates emission in fan beams \citep{wpz+14,dr15}. The fan extends to much larger distances from the pulsar surface and hence emission at high impact angles is possible. As such high impact angles are rarely observed, \citet{wpz+14} surmised that the flux density would decrease as the impact angle increases. In this picture therefore, distant pulsars would be seen only at low impact angles unlike their nearby counterparts.

Disentangling differences in morphology between the profiles of MC pulsars and their Galactic counterparts is made more difficult by the severe selection effects in surveys for MC pulsars. Modern search techniques (whether Fourier transform or fast-folding based) are biased in favour of pulsars with narrow pulse profiles and against those with wider profiles \citep{vhkr17}. As we only detect 1 pulsar per 1000 in the LMC/SMC system \citep{rl10,ttm+20} then it comes as no surprise to see that the majority have narrow profiles. Furthermore, the lack of pulsars with $\dot{E} > 10^{34}$~erg\,s$^{-1}$ in the MCs (apart from PSR~J0540--6919 which was not discovered in a blind survey), is likely because these pulsars tend to have wider profiles than their low $\dot{E}$ counterparts (e.g. \citealt{jw06}).

One potential approach is to compare the pulse widths of the MC pulsars with the Galactic counterparts also detected near the survey limit. If survey biases dominate, then we would expect that the pulsars would, on average, have smaller widths than expected. Conversely, we might expect the bright Galactic pulsars to have wider profiles on average. The MC pulsars typically have low flux densities ($< 0.04$~mJy) and are found near the survey detection limit. As a comparison set, we choose Galactic pulsars detected in the High Time Resolution (HTRU) survey with the Parkes telescope \citep{kjv+10} with flux densities $< 0.25$~mJy. There are 225 such pulsars, of which only 11 (4.8\%) have $\dot{E}>10^{34}$~erg\,s$^{-1}$. This would imply that high $\dot{E}$ pulsars are only marginally under-represented in the MC sample. For the bright pulsars we choose those with $S>1.5$~mJy from the HTRU survey. There are 203 such pulsars of which 24\% have $\dot{E}>10^{34}$~erg\,s$^{-1}$.

Recently, \citet{poss21} measured the profile width at 10\% of the peak for many hundreds of pulsars and plotted these values versus the pulsar's spin period, $P$. For any given $P$ a large scatter in the pulse width is seen but \citet{poss21} derived a best fit to the data and found that
\begin{equation}
    W_e = 11.9 \,\, P^{-0.63}
\end{equation}
where $W_e$ is the expected width in degrees for $P$ in seconds.
They also measured values of the width, $W_m$, for 68 pulsars out of the 225 low flux density pulsars selected above and 169 of the bright sample. We can estimate values of $W_m$ for 23 of the MC sample. We consider that $W_e = W_m$ if the difference is less than 2\degr, and otherwise crudely divide the sample into $W_e>W_m$ and $W_e<W_m$. Results are shown in Table~\ref{tab_width}. The LMC/SMC sample show a very similar width distribution to the low flux-density HTRU sample in that they favour narrow observed widths, whereas the bright pulsars from HTRU show evidence for wider pulsars than expectations. This clearly shows that the reason the MC pulsars have narrow profiles is due to the selection effects in pulsar surveys.
\begin{table}
\caption{A comparison of measured widths ($W_m$) and expected widths ($W_e$) for four samples defined in the text. N denotes the total number in the sample.}
\label{tab_width}
\begin{center}
    \begin{tabular}{lrccc}
    \hline
    \hline
Sample &  N & $W_m<W_e$ & $W_m=W_e$ & $W_m>W_e$ \\
\hline
LMC/SMC  &  23 & 48\% & 22\% & 30\% \\
HTRU, $S<0.25$~mJy & 68 & 50\% & 24\% & 26\% \\
HTRU, $S>1.5$~mJy & 169 & 36\% & 16\% & 49\%\\
HTRU, $L>125$~mJy\,kpc$^2$ & 49 & 28\% & 10\% & 61\%\\
\hline
    \end{tabular}
\end{center}
\end{table}

On the other hand, if distant (high luminosity) pulsars have low impact angle compared to nearby pulsars then their pulse width distributions should be different. \citet{rcl+13} and \citet{mfl+06} have shown that the luminosity function above 125~mJy\,kpc$^2$ is the same for Galactic and MC pulsars. In a similar fashion to above, we construct a high luminosity sample from the HTRU survey and take the pulse widths from \citet{poss21}. Results are shown in the bottom row of Table~\ref{tab_width}. We see that the profiles of the high luminosity Galactic sample are significantly wider than those of the Magellanic sample. This again demonstrates that the morphologies of the Magellanic pulsars are dominated by selection effects towards narrow profiles.

\begin{figure}
\begin{center}
\begin{tabular}{c}
\includegraphics[width=8cm,angle=0]{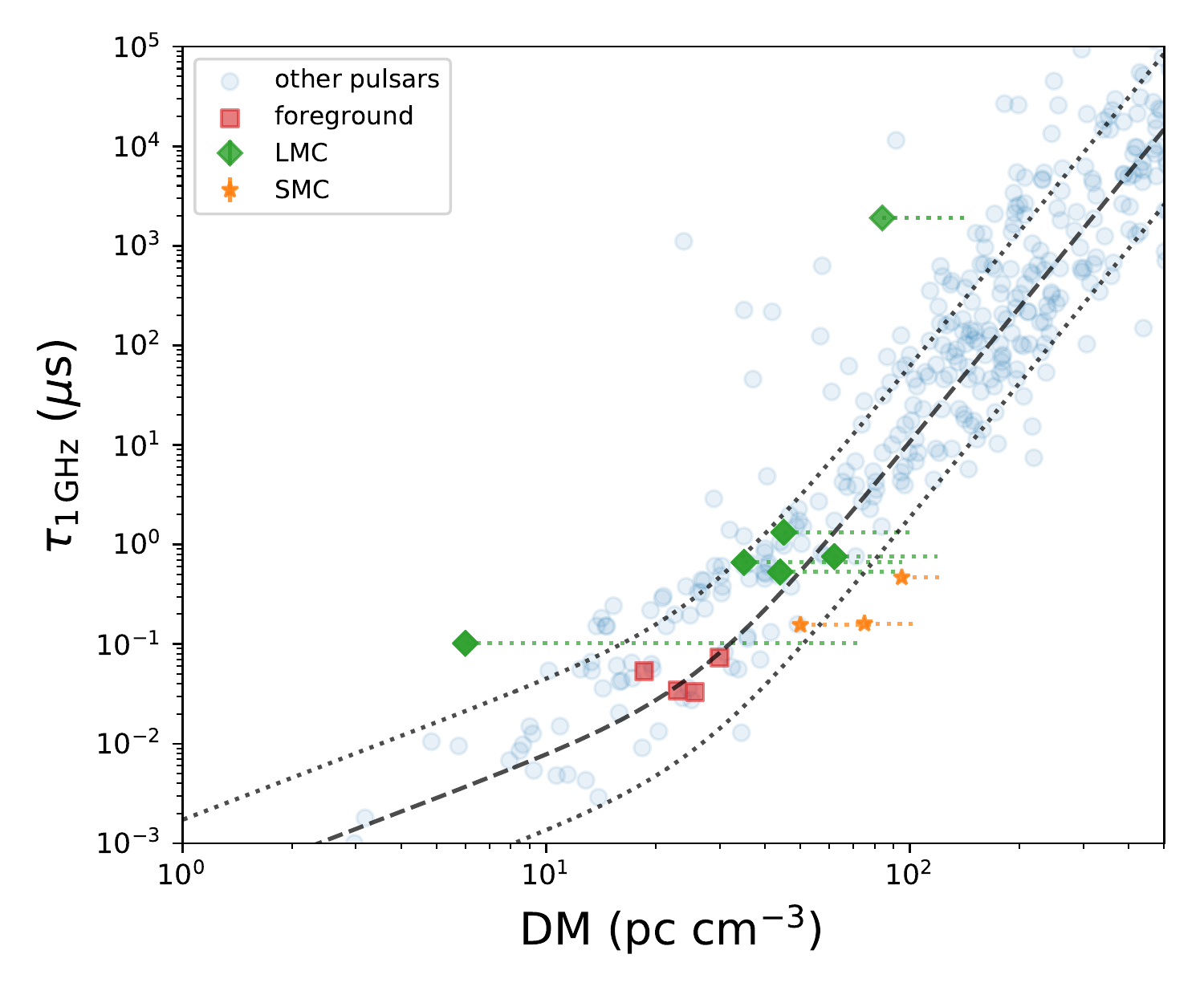} \\
\end{tabular}
\end{center}
\caption{Dispersion Measure versus scattering time for selected pulsars. Green diamonds and orange stars denote LMC and SMC pulsars respectively, red squares denote foreground pulsars and blue dots other pulsars with measurements from the literature.  For each MC pulsar, the horizontal dotted lines connect the DM within the MCs to the total DM. The dash and dotted black lines show the fit of \citet{cws+16} and associated error estimation.}
\label{figscint}
\end{figure}
There remains a possibility that MC pulsars are inherently different to Galactic pulsars as, for example, the metallicity of the MCs is different to that in the Milky Way. This could manifest itself as a difference in initial spin and magnetic properties of the MC pulsars. With only a small number of pulsars at our disposal this is difficult to test, albeit attempted by \citet{ttm+20} for the SMC.

\subsection{Scintillation}
We were able to measure the diffractive scintillation bandwidth, $\Delta\nu_d$, for four of the foreground pulsars and eight of the brighter MC pulsars. For the low DM foreground pulsars, $\Delta\nu_d$ ranges from 2 to 5~MHz. For the MC pulsars, $\Delta\nu_d$ ranges from 0.1 to 1.6~MHz. In addition we include the measurement of scattering in PSR~J0540--6919 by \citet{gsa+21}. There is a relationship between the scattering time (and therefore $\Delta\nu_d$) and DM for Galactic pulsars as given most recently by \citet{kmn+15} and \citet{cws+16}. If we assume the scintillation in the MC pulsars is caused by the interstellar medium within the MCs themselves we can place these pulsars on the same relationship by using the DMs given in Tables~\ref{tab_blmc} and \ref{tab_bsmc}.

Figure~\ref{figscint} shows scattering time versus DM for Galactic pulsars (from the literature) and the MC pulsars from this work in addition to the relationship of \citet{cws+16}. The foreground pulsars fall nicely on the trend. For the LMC pulsars, PSR~J0540--6919 has very large scattering for its DM. This scattering almost certainly arises from the pulsar's parent supernova remnant in the same way as the Vela pulsar is heavily scattered by the Gum nebula. The other LMC pulsars fall on the trend, apart from PSR~J0555--7056 for which it is possible that the DM contribution of the LMC has been underestimated by \citet{ymw17}. The three SMC pulsars appear slightly under-scattered, even though we expect the scattering to be a factor two higher than the standard Galactic case for a pulsar with a screen located at half the pulsar's distance. Again, the DM contribution from the Galaxy and its halo may be underestimated (and hence the SMC contribution overestimated) by \citet{ymw17} and a discussion of this possibility can be found in \citet{pfd21}. The flux density variations of more than a factor of 10 in PSR~J0045--7319 could be from large-scale refractive effects, or the presence of an achromatic second screen near the pulsar. It is possible that the scintillation may still arise from the Milky Way rather than the MCs, which a single measurement of $\nu_d$ does not rule out. A measurement of annual variation in the scintillation timescale, or an angular broadening measurement would distinguish between the two cases.

\section{Summary}
We have provided polarization profiles for 14 pulsars in the LMC and SMC and updates on their DMs and RMs. The inferred values of $B_{||}$ are in line with expectations for the LMC and SMC, although local conditions such as those around PSR~J0540--6919 are also important. We show that the MC pulsars are narrower than expected from the period-width relationship of \citet{poss21} and demonstrate that this is likely a selection effect in the pulsar surveys that detected them, rather than an intrinsic property of the pulsars themselves. The scintillation parameters for the pulsars in the LMC are broadly in line with expectations apart from PSR~J0540--6919 which has enhanced scattering from its parent supernova remnant. In the SMC, the relatively low scattering may imply that the DM contribution from the Galactic halo is underestimated. Further observations are required to determine the location of the scattering screens.

Drawing significant conclusions from the data is hampered by the small number of known pulsars in the Magellanic system. In the near future, MeerKAT will survey the Magellanic Clouds for pulsars to a much greater depth than the previous Parkes surveys \citep{trapum}. This should allow the ideas put forward in Sections~4.2 and 4.3 to be explored in more detail and should fill in the pulsar contribution to Figure~\ref{figbig}. Finally, many more RMs to extra-galactic sources seen through the MCs should be made possible through the surveys on the ASKAP telescope \citep{jtb+08,glt+10}.

\section*{Data Availability}
The data underlying this article will be shared on reasonable request to the corresponding author.

\section*{Acknowledgements}
We thank G.~Heald and D.~Schnitzeler for useful discussions.
The MeerKAT telescope is operated by the South African Radio Astronomy Observatory, which is a facility of the National Research Foundation, an agency of the Department of Science and Innovation. MeerTime data is housed and processed on the OzSTAR supercomputer at Swinburne University of Technology with the support of ADACS and the gravitational wave data centre via AAL. RMS acknowledges support through Australian Research Council Future Fellowship FT190100155. VVK acknowledges support from the Max-Planck Society.

\bibliographystyle{mnras}
\bibliography{mcs} % if your bibtex file is called example.bib

\begin{thebibliography}{}
\makeatletter
\relax
\def\mn@urlcharsother{\let\do\@makeother \do\$\do\&\do\#\do\^\do\_\do\%\do\~}
\def\mn@doi{\begingroup\mn@urlcharsother \@ifnextchar [ {\mn@doi@}
  {\mn@doi@[]}}
\def\mn@doi@[#1]#2{\def\@tempa{#1}\ifx\@tempa\@empty \href
  {http://dx.doi.org/#2} {doi:#2}\else \href {http://dx.doi.org/#2} {#1}\fi
  \endgroup}
\def\mn@eprint#1#2{\mn@eprint@#1:#2::\@nil}
\def\mn@eprint@arXiv#1{\href {http://arxiv.org/abs/#1} {{\tt arXiv:#1}}}
\def\mn@eprint@dblp#1{\href {http://dblp.uni-trier.de/rec/bibtex/#1.xml}
  {dblp:#1}}
\def\mn@eprint@#1:#2:#3:#4\@nil{\def\@tempa {#1}\def\@tempb {#2}\def\@tempc
  {#3}\ifx \@tempc \@empty \let \@tempc \@tempb \let \@tempb \@tempa \fi \ifx
  \@tempb \@empty \def\@tempb {arXiv}\fi \@ifundefined
  {mn@eprint@\@tempb}{\@tempb:\@tempc}{\expandafter \expandafter \csname
  mn@eprint@\@tempb\endcsname \expandafter{\@tempc}}}

\bibitem[\protect\citeauthoryear{{Bailes} et~al.,}{{Bailes}
  et~al.}{2020}]{mtime}
{Bailes} M.,  et~al., 2020, PASA, 37, e028

\bibitem[\protect\citeauthoryear{{Bell}, {Bessell}, {Stappers}, {Bailes}  \&
  {Kaspi}}{{Bell} et~al.}{1995}]{bbs+95}
{Bell} J.~F.,  {Bessell} M.~S.,  {Stappers} B.~W.,  {Bailes} M.,   {Kaspi}
  V.~M.,  1995, ApJ, 447, L117

\bibitem[\protect\citeauthoryear{{Brantseg}, {McEntaffer}, {Bozzetto},
  {Filipovic}  \& {Grieves}}{{Brantseg} et~al.}{2014}]{bmb+14}
{Brantseg} T.,  {McEntaffer} R.~L.,  {Bozzetto} L.~M.,  {Filipovic} M.,
  {Grieves} N.,  2014, ApJ, 780, 50

\bibitem[\protect\citeauthoryear{{Brisken}, {Macquart}, {Gao}, {Rickett},
  {Coles}, {Deller}, {Tingay}  \& {West}}{{Brisken} et~al.}{2010}]{bmg+10}
{Brisken} W.~F.,  {Macquart} J.~P.,  {Gao} J.~J.,  {Rickett} B.~J.,  {Coles}
  W.~A.,  {Deller} A.~T.,  {Tingay} S.~J.,   {West} C.~J.,  2010, ApJ, 708, 232

\bibitem[\protect\citeauthoryear{{Cordes}, {Wharton}, {Spitler}, {Chatterjee}
  \& {Wasserman}}{{Cordes} et~al.}{2016}]{cws+16}
{Cordes} J.~M.,  {Wharton} R.~S.,  {Spitler} L.~G.,  {Chatterjee} S.,
  {Wasserman} I.,  2016, arXiv e-prints, p. arXiv:1605.05890

\bibitem[\protect\citeauthoryear{{Crawford}, {Manchester}  \&
  {Kaspi}}{{Crawford} et~al.}{2001a}]{cmk01}
{Crawford} F.,  {Manchester} R.~N.,   {Kaspi} V.~M.,  2001a, AJ, 122, 2001

\bibitem[\protect\citeauthoryear{{Crawford}, {Kaspi}, {Manchester}, {Lyne},
  {Camilo}  \& {D'Amico}}{{Crawford} et~al.}{2001b}]{ckm+01}
{Crawford} F.,  {Kaspi} V.~M.,  {Manchester} R.~N.,  {Lyne} A.~G.,  {Camilo}
  F.,   {D'Amico} N.,  2001b, ApJ, 553, 367

\bibitem[\protect\citeauthoryear{{Dickel}, {Gruendl}, {McIntyre}  \&
  {Amy}}{{Dickel} et~al.}{2010}]{dgma10}
{Dickel} J.~R.,  {Gruendl} R.~A.,  {McIntyre} V.~J.,   {Amy} S.~W.,  2010, AJ,
  140, 1511

\bibitem[\protect\citeauthoryear{{Dyks} \& {Rudak}}{{Dyks} \&
  {Rudak}}{2015}]{dr15}
{Dyks} J.,  {Rudak} B.,  2015, MNRAS, 446, 2505

\bibitem[\protect\citeauthoryear{{Everett} \& {Weisberg}}{{Everett} \&
  {Weisberg}}{2001}]{ew01}
{Everett} J.~E.,  {Weisberg} J.~M.,  2001, ApJ, 553, 341

\bibitem[\protect\citeauthoryear{{Gaensler} \& {POSSUM
  Collaboration}}{{Gaensler} \& {POSSUM Collaboration}}{2010}]{glt+10}
{Gaensler} B.~M.,  {POSSUM Collaboration} 2010, in American Astronomical
  Society Meeting Abstracts \#215. p. 470.13

\bibitem[\protect\citeauthoryear{{Gaensler}, {Haverkorn}, {Staveley-Smith},
  {Dickey}, {McClure-Griffiths}, {Dickel}  \& {Wolleben}}{{Gaensler}
  et~al.}{2005}]{ghs+05}
{Gaensler} B.~M.,  {Haverkorn} M.,  {Staveley-Smith} L.,  {Dickey} J.~M.,
  {McClure-Griffiths} N.~M.,  {Dickel} J.~R.,   {Wolleben} M.,  2005, Science,
  307, 1610

\bibitem[\protect\citeauthoryear{{Geyer} et~al.,}{{Geyer}
  et~al.}{2021}]{gsa+21}
{Geyer} M.,  et~al., 2021, MNRAS, 505, 4468

\bibitem[\protect\citeauthoryear{{Graczyk} et~al.,}{{Graczyk}
  et~al.}{2020}]{gpt+20}
{Graczyk} D.,  et~al., 2020, ApJ, 904, 13

\bibitem[\protect\citeauthoryear{{Han}, {Manchester}, {van Straten}  \&
  {Demorest}}{{Han} et~al.}{2018}]{hmvd18}
{Han} J.~L.,  {Manchester} R.~N.,  {van Straten} W.,   {Demorest} P.,  2018,
  ApJS, 234, 11

\bibitem[\protect\citeauthoryear{{Hobbs}, {Edwards}  \& {Manchester}}{{Hobbs}
  et~al.}{2006}]{hem06}
{Hobbs} G.~B.,  {Edwards} R.~T.,   {Manchester} R.~N.,  2006, MNRAS, 369, 655

\bibitem[\protect\citeauthoryear{{Hotan}, {van Straten}  \&
  {Manchester}}{{Hotan} et~al.}{2004}]{hvm04}
{Hotan} A.~W.,  {van Straten} W.,   {Manchester} R.~N.,  2004, PASA, 21, 302

\bibitem[\protect\citeauthoryear{{Johnston} \& {Kerr}}{{Johnston} \&
  {Kerr}}{2018}]{jk18}
{Johnston} S.,  {Kerr} M.,  2018, MNRAS, 474, 4629

\bibitem[\protect\citeauthoryear{{Johnston} \& {Romani}}{{Johnston} \&
  {Romani}}{2003}]{jr03}
{Johnston} S.,  {Romani} R.~W.,  2003, ApJ, 590, L95

\bibitem[\protect\citeauthoryear{{Johnston} \& {Weisberg}}{{Johnston} \&
  {Weisberg}}{2006}]{jw06}
{Johnston} S.,  {Weisberg} J.~M.,  2006, MNRAS, 368, 1856

\bibitem[\protect\citeauthoryear{{Johnston}, {Nicastro}  \&
  {Koribalski}}{{Johnston} et~al.}{1998}]{jnk98}
{Johnston} S.,  {Nicastro} L.,   {Koribalski} B.,  1998, MNRAS, 297, 108

\bibitem[\protect\citeauthoryear{{Johnston} et~al.,}{{Johnston}
  et~al.}{2008}]{jtb+08}
{Johnston} S.,  et~al., 2008, Experimental Astronomy, 22, 151

\bibitem[\protect\citeauthoryear{{Johnston} et~al.,}{{Johnston}
  et~al.}{2020}]{tpa20}
{Johnston} S.,  et~al., 2020, MNRAS, 493, 3608

\bibitem[\protect\citeauthoryear{{Karastergiou} \& {Johnston}}{{Karastergiou}
  \& {Johnston}}{2007}]{kj07}
{Karastergiou} A.,  {Johnston} S.,  2007, MNRAS, 380, 1678

\bibitem[\protect\citeauthoryear{{Kaspi}, {Johnston}, {Bell}, {Manchester},
  {Bailes}, {Bessell}, {Lyne}  \& {D'Amico}}{{Kaspi} et~al.}{1994}]{kjb+94}
{Kaspi} V.~M.,  {Johnston} S.,  {Bell} J.~F.,  {Manchester} R.~N.,  {Bailes}
  M.,  {Bessell} M.,  {Lyne} A.~G.,   {D'Amico} N.,  1994, ApJ, 423, L43

\bibitem[\protect\citeauthoryear{{Keith} et~al.,}{{Keith}
  et~al.}{2010}]{kjv+10}
{Keith} M.~J.,  et~al., 2010, MNRAS, 409, 619

\bibitem[\protect\citeauthoryear{{Kerr}, {Coles}, {Ward}, {Johnston}, {Tuntsov}
   \& {Shannon}}{{Kerr} et~al.}{2018}]{kcw+18}
{Kerr} M.,  {Coles} W.~A.,  {Ward} C.~A.,  {Johnston} S.,  {Tuntsov} A.~V.,
  {Shannon} R.~M.,  2018, MNRAS, 474, 4637

\bibitem[\protect\citeauthoryear{{Kim}, {Staveley-Smith}, {Dopita}, {Freeman},
  {Sault}, {Kesteven}  \& {McConnell}}{{Kim} et~al.}{1998}]{ksd+98}
{Kim} S.,  {Staveley-Smith} L.,  {Dopita} M.~A.,  {Freeman} K.~C.,  {Sault}
  R.~J.,  {Kesteven} M.~J.,   {McConnell} D.,  1998, ApJ, 503, 674

\bibitem[\protect\citeauthoryear{{Krishnakumar}, {Mitra}, {Naidu}, {Joshi}  \&
  {Manoharan}}{{Krishnakumar} et~al.}{2015}]{kmn+15}
{Krishnakumar} M.~A.,  {Mitra} D.,  {Naidu} A.,  {Joshi} B.~C.,   {Manoharan}
  P.~K.,  2015, ApJ, 804, 23

\bibitem[\protect\citeauthoryear{{Kumamoto} et~al.,}{{Kumamoto}
  et~al.}{2021}]{kdj+21}
{Kumamoto} H.,  et~al., 2021, MNRAS, 501, 4490

\bibitem[\protect\citeauthoryear{{Manchester}, {Mar}, {Lyne}, {Kaspi}  \&
  {Johnston}}{{Manchester} et~al.}{1993}]{mml+93}
{Manchester} R.~N.,  {Mar} D.~P.,  {Lyne} A.~G.,  {Kaspi} V.~M.,   {Johnston}
  S.,  1993, ApJ, 403, L29

\bibitem[\protect\citeauthoryear{{Manchester}, {Fan}, {Lyne}, {Kaspi}  \&
  {Crawford}}{{Manchester} et~al.}{2006}]{mfl+06}
{Manchester} R.~N.,  {Fan} G.,  {Lyne} A.~G.,  {Kaspi} V.~M.,   {Crawford} F.,
  2006, ApJ, 649, 235

\bibitem[\protect\citeauthoryear{{Mao}, {Gaensler}, {Stanimirovi{\'c}},
  {Haverkorn}, {McClure-Griffiths}, {Staveley-Smith}  \& {Dickey}}{{Mao}
  et~al.}{2008}]{mgs+08}
{Mao} S.~A.,  {Gaensler} B.~M.,  {Stanimirovi{\'c}} S.,  {Haverkorn} M.,
  {McClure-Griffiths} N.~M.,  {Staveley-Smith} L.,   {Dickey} J.~M.,  2008,
  ApJ, 688, 1029

\bibitem[\protect\citeauthoryear{{Mao} et~al.,}{{Mao} et~al.}{2012}]{mmg+12}
{Mao} S.~A.,  et~al., 2012, ApJ, 759, 25

\bibitem[\protect\citeauthoryear{{McConnell}, {McCulloch}, {Hamilton}, {Ables},
  {Hall}, {Jacka}  \& {Hunt}}{{McConnell} et~al.}{1991}]{mmh+91}
{McConnell} D.,  {McCulloch} P.~M.,  {Hamilton} P.~A.,  {Ables} J.~G.,  {Hall}
  P.~J.,  {Jacka} C.~E.,   {Hunt} A.~J.,  1991, MNRAS, 249, 654

\bibitem[\protect\citeauthoryear{{McCulloch}, {Hamilton}, {Ables}  \&
  {Hunt}}{{McCulloch} et~al.}{1983}]{mhah83}
{McCulloch} P.~M.,  {Hamilton} P.~A.,  {Ables} J.~G.,   {Hunt} A.~J.,  1983,
  Nature, 303, 307

\bibitem[\protect\citeauthoryear{{Noutsos}, {Johnston}, {Kramer}  \&
  {Karastergiou}}{{Noutsos} et~al.}{2008}]{njkk08}
{Noutsos} A.,  {Johnston} S.,  {Kramer} M.,   {Karastergiou} A.,  2008, MNRAS,
  386, 1881

\bibitem[\protect\citeauthoryear{{Parthasarathy} et~al.,}{{Parthasarathy}
  et~al.}{2021}]{pbs21}
{Parthasarathy} A.,  et~al., 2021, MNRAS, 502, 407

\bibitem[\protect\citeauthoryear{{Pietrzy{\'n}ski} et~al.,}{{Pietrzy{\'n}ski}
  et~al.}{2019}]{pgg+19}
{Pietrzy{\'n}ski} G.,  et~al., 2019, Nature, 567, 200

\bibitem[\protect\citeauthoryear{{Posselt} et~al.,}{{Posselt}
  et~al.}{2021}]{poss21}
{Posselt} B.,  et~al., 2021, MNRAS. In Press.

\bibitem[\protect\citeauthoryear{{Price}, {Flynn}  \& {Deller}}{{Price}
  et~al.}{2021}]{pfd21}
{Price} D.~C.,  {Flynn} C.,   {Deller} A.,  2021, PASA, 38, e038

\bibitem[\protect\citeauthoryear{{Rankin}}{{Rankin}}{1990}]{ran90}
{Rankin} J.~M.,  1990, ApJ, 352, 247

\bibitem[\protect\citeauthoryear{{Ridley} \& {Lorimer}}{{Ridley} \&
  {Lorimer}}{2010}]{rl10}
{Ridley} J.~P.,  {Lorimer} D.~R.,  2010, MNRAS, 406, L80

\bibitem[\protect\citeauthoryear{{Ridley}, {Crawford}, {Lorimer}, {Bailey},
  {Madden}, {Anella}  \& {Chennamangalam}}{{Ridley} et~al.}{2013}]{rcl+13}
{Ridley} J.~P.,  {Crawford} F.,  {Lorimer} D.~R.,  {Bailey} S.~R.,  {Madden}
  J.~H.,  {Anella} R.,   {Chennamangalam} J.,  2013, MNRAS, 433, 138

\bibitem[\protect\citeauthoryear{{Schnitzeler}}{{Schnitzeler}}{2012}]{sch12}
{Schnitzeler} D.~H.~F.~M.,  2012, MNRAS, 427, 664

\bibitem[\protect\citeauthoryear{{Serylak} et~al.,}{{Serylak}
  et~al.}{2021}]{sjk+21}
{Serylak} M.,  et~al., 2021, MNRAS, 505, 4483

\bibitem[\protect\citeauthoryear{{Sobey} et~al.,}{{Sobey}
  et~al.}{2019}]{sbg+19}
{Sobey} C.,  et~al., 2019, MNRAS, 484, 3646

\bibitem[\protect\citeauthoryear{{Stappers} \& {Kramer}}{{Stappers} \&
  {Kramer}}{2016}]{trapum}
{Stappers} B.,  {Kramer} M.,  2016, in MeerKAT Science: On the Pathway to the
  SKA. p.~9

\bibitem[\protect\citeauthoryear{{Szary}, {Zhang}, {Melikidze}, {Gil}  \&
  {Xu}}{{Szary} et~al.}{2014}]{szm+14}
{Szary} A.,  {Zhang} B.,  {Melikidze} G.~I.,  {Gil} J.,   {Xu} R.-X.,  2014,
  ApJ, 784, 59

\bibitem[\protect\citeauthoryear{{Titus} et~al.,}{{Titus}
  et~al.}{2019}]{tsm+19}
{Titus} N.,  et~al., 2019, MNRAS, 487, 4332

\bibitem[\protect\citeauthoryear{{Titus}, {Toonen}, {McBride}, {Stappers},
  {Buckley}  \& {Levin}}{{Titus} et~al.}{2020}]{ttm+20}
{Titus} N.,  {Toonen} S.,  {McBride} V.~A.,  {Stappers} B.~W.,  {Buckley}
  D.~A.~H.,   {Levin} L.,  2020, MNRAS, 494, 500

\bibitem[\protect\citeauthoryear{{Wang} et~al.,}{{Wang} et~al.}{2014}]{wpz+14}
{Wang} H.~G.,  et~al., 2014, ApJ, 789, 73

\bibitem[\protect\citeauthoryear{{Yao}, {Manchester}  \& {Wang}}{{Yao}
  et~al.}{2017}]{ymw17}
{Yao} J.~M.,  {Manchester} R.~N.,   {Wang} N.,  2017, ApJ, 835, 29

\bibitem[\protect\citeauthoryear{{van Heerden}, {Karastergiou}  \&
  {Roberts}}{{van Heerden} et~al.}{2017}]{vhkr17}
{van Heerden} E.,  {Karastergiou} A.,   {Roberts} S.~J.,  2017, MNRAS, 467,
  1661

\bibitem[\protect\citeauthoryear{{van Straten}, {Manchester}, {Johnston}  \&
  {Reynolds}}{{van Straten} et~al.}{2010}]{vmjr10}
{van Straten} W.,  {Manchester} R.~N.,  {Johnston} S.,   {Reynolds} J.~E.,
  2010, PASA, 27, 104

\makeatother
\end{thebibliography}
%%%%%%%%%%%%%%%%%%%%%%%%%%%%%%%%%%%%%%%%%%%%%%%%%%

% Don't change these lines
\bsp	% typesetting comment
\label{lastpage}
\end{document}